\documentclass[onecolumn]{aastex631}

\newcommand{\fancyE}{\mathcal{E}}
\newcommand{\rom}[1]{\uppercase\expandafter{\romannumeral #1\relax}}

\usepackage{amsmath}

\graphicspath{{./}{figures/}}
\begin{document}  


\title{The Effect of the Parametric Decay Instability on the Morphology of Coronal Type \rom{3} Radio Bursts}

\correspondingauthor{Chaitanya Prasad Sishtla}
\email{chaitanya.sishtla@helsinki.fi}

\author[0000-0003-4236-768X]{Chaitanya Prasad Sishtla}
\affiliation{Department of Physics, University of Helsinki,
          Helsinki, Finland}

\author[0000-0002-0606-7172]{Immanuel Christopher Jebaraj}
\affiliation{Department of Physics \& Astronomy, University of Turku,
          Turku, Finland}

\author[0000-0003-1175-7124]{Jens Pomoell}
\affiliation{Department of Physics, University of Helsinki,
          Helsinki, Finland}

\author[0000-0001-5731-8173]{Norbert Magyar}
\affiliation{Centre for mathematical Plasma Astrophysics, Department of Mathematics, KU Leuven, Leuven, Belgium}


\author[0000-0002-1573-7457]{Marc Pulupa}
\affil{Space Sciences Laboratory, University of California, Berkeley, CA 94720-7450, USA}

\author[0000-0002-4489-8073]{Emilia Kilpua}
\affiliation{Department of Physics, University of Helsinki,
          Helsinki, Finland}

\author[0000-0002-1989-3596]{Stuart D. Bale}
\affil{Space Sciences Laboratory, University of California, Berkeley, CA 94720-7450, USA}
\affil{Physics Department, University of California, Berkeley, CA 94720-7300, USA}



\begin{abstract}
The nonlinear evolution of Alfv\'en waves in the solar corona leads to the generation of Alfv\'enic turbulence. This description of the Alfv\'en waves involves parametric instabilities where the parent wave decays into slow mode waves giving rise to density fluctuations. These density fluctuations, in turn, play a crucial role in the modulation of the dynamic spectrum of type \rom{3} radio bursts, which are observed at the fundamental of local plasma frequency and are sensitive to the local density. During observations of such radio bursts, fine structures are detected across different temporal ranges. In this study, we examine density fluctuations generated through the parametric decay instability (PDI) of Alfv\'en waves as a mechanism to generate striations in the dynamic spectrum of type \rom{3} radio bursts using magnetohydrodynamic simulations of the solar corona. An Alfv\'en wave is injected into the quiet solar wind by perturbing the transverse magnetic field and velocity components which subsequently undergo the PDI instability. The type \rom{3} burst is modelled as a fast-moving radiation source that samples the background solar wind as it propagates to emit radio waves. We find the simulated dynamic spectrum to contain striations directly affected by the multi-scale density fluctuations in the wind.
\end{abstract}


%
\section{Introduction}\label{sec:introduction}
The expanding solar wind contains fluctuations in the velocity and magnetic field that are primarily transverse to the mean field and have characteristics of~\cite{kolmogorov1991local} turbulence. In-situ studies in the heliosphere show that most of the wave power in solar wind fluctuations result from Alfv\'enic waves, while non-Alfv\'enic waves have a minor contribution~\citep{higdon1984density, bruno2013solar, chen2016recent}. This allows modelling the ubiquitous solar wind fluctuations as Alfv\'en waves~\citep{belcher1971large, dAmicis2015}. Subsequently, the heating and acceleration of the solar wind is explained via the cascade of large-wavelength Alfv\'en waves leading to Alfv\'enic turbulence~\citep{coleman1968turbulence, goldreich1995toward}. The Alfv\'enic fluctuations are generated, e.g., by the photospheric convective motions where the solar magnetic field footpoints are anchored ~\citep{cranmer2005generation}.  
As these waves propagate outward from the Sun, they can decay into slow modes through modulational instabilities~\citep{lashmore1976modulational, sakai1983modulational}, and generate density variations through nonlinear wave-wave interactions between fast, slow, and Alfv\'en waves~\citep{nakariakov2000nonlinear, chandran2005weak}. In particular, Alfv\'en waves can couple with background gradients in the background magnetic field to decay into a slow wave and a reflected (sunward) Alfv\'en wave through the parametric decay instability (PDI)~\citep{galeev1963stability,sagdeev1969nonlinear, Derby1978, hoshino1989time}. The PDI instability has been observed both in the solar wind~\citep{bowen2018density} and in the laboratory~\citep{dorfman2016labPDI}, and its importance in generating counter-propagating waves leading to Alfv\'enic turbulence~\citep{Usmanov2000, SuzukiInutsuka2006} is a topic of continued interest~\citep{chandran2018parametric, shoda2018frequency, sishtla2022flux}. The relevance of the PDI for Alfv\'en wave dynamics is of particular significance in the solar corona due to the low-beta plasma present in this region. Hinode observations~\citep{Pontieu2007} characterise the Alfv\'en waves injected into the solar corona to have wave periods between $100-500$ s, and amplitudes between $20-50$ km s$^{-1}$. As the waves further propagate in the solar corona, the low plasma beta environment~\citep{Iwai2014, Bourdin2017} increases substantially the decay rates of the pump Alfv\'en wave due to the lower threshold of the PDI in a low beta plasma~\citep{goldstein1978instability}. In this context, the PDI can contribute to coronal heating and the observed spectrum of cross-helicity in the solar wind~\citep{Inchester1990PDI, delZana2001PDI} by generating counter-propagating Alfv\'en waves. The present study underscores the significance of PDI by demonstrating that it has a direct effect on other phenomena as well, in particular the formation of striations and other fine structures within Type \rom{3} solar radio bursts.



\begin{figure*}
    \centering
    \includegraphics[width=\linewidth]{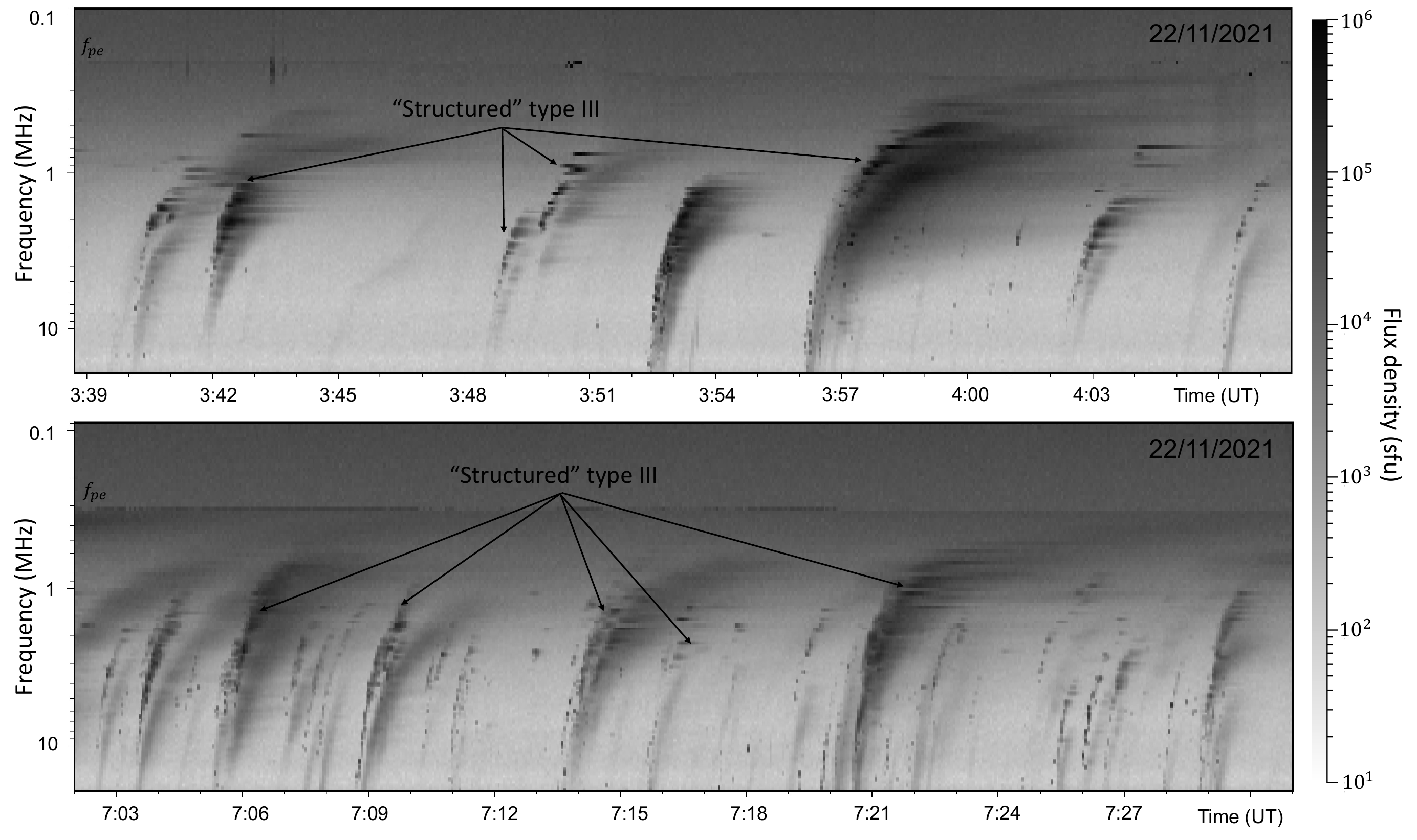}
    \caption{Example dynamic radio spectra produced from the measurements of FIELDS/RFS onboard PSP during encounter 9 (22/11/2021). The probe was approximately 17 $R_{\odot}$ during the time intervals at which these measurements were made. A few examples of the ``structured" type III radio bursts are indicated in the top panel. The local plasma frequency at the spacecraft ($f_{pe}$) is also marked in both spectra. }
    \label{fig:type3_example}
\end{figure*}

Type \rom{3} radio bursts are signatures of suprathermal electron beams propagating along open magnetic field lines~\citep[e.g.,][]{Book1985SolarRadio, Pick08}, with typical beam speeds for $\approx 30$ keV electrons between $0.1c-0.4c$ (where $c$ is the speed of light in vacuum)~\citep{Zhang18}. Due to their near-relativistic velocities, electromagnetic emissions from these beams can be used to probe the nature of the background plasma through which the beams propagate. Type \rom{3} radio bursts are the most commonly observed and the most intense radio bursts of solar origin. Two example intervals of the Radio Frequency Spectrometer \citep[RFS;][]{Pulupa2017} which is part of the FIELDS instrument suite \citep[][]{Bale16} onboard the Parker Solar Probe \citep[PSP;][]{Fox2016} is shown in Figure \ref{fig:type3_example}. The measurements were made during PSP's 9th close encounter and show a number of type \rom{3} radio bursts. Type \rom{3} bursts as the ones recorded in Figure \ref{fig:type3_example}, appear as fast drifting bursts over a wide range of frequencies~\citep[e.g.,][]{Suzuki1985}. The most widely accepted generation mechanism is coherent emission and was first proposed by~\cite{Ginzburg1958}. This so-called ``plasma emission mechanism'' is a two step process. The first step involves an electron beam which becomes unstable due to time-of-flight effects and generates Langmuir waves. In the second step, the Langmuir waves linearly or non-linearly convert part of their energy into electromagnetic waves at radio frequencies. As a natural consequence of the generation mechanism, emission is predicted at the fundamental plasma frequency $f_{pe}$, and its harmonics $nf_{pe}$ ~\citep[where n = 2,3,\dots][]{Robinson1998Part1, Robinson1998Part2, Robinson1998Part3}. \cite{Melrose1980} expanded upon the original idea of non-linear processes which included scattering off ion-sound waves for the generation of fundamental radiation. However, later models demonstrated that a linear conversion was preferable in a randomly inhomogeneous plasma \citep[][]{Hinkel92, Kim2007, Voshchep15a, Voshchep15b, Krasnoselskikh19}. Other competing mechanisms for the generation of electromagnetic radiation through beam-driven Langmuir include, electrostatic decay \citep[][]{Cairns85, Cairns88, Robinson1994}, electromagnetic decay \citep[][]{Cairns1987, Robinson1993}, quasi-mode processes \citep[][]{Yoon94}, non-zero pitch angle electron beam driven radiation \citep[][]{Tsiklauri2011, Schmitz2013}, and antenna radiation \citep[][]{Papadopoulos78, Malaspina2010, Malaspina12}. Simulating the local evolution of the beam-plasma system and the emission of electromagnetic radiation self-consistently requires kinetic \citep[][]{Thurgood2015, Krafft22a} or other approaches used to treat Langmuir turbulence, such as using the Zakharov-Hamiltonian set of equations \citep[][]{Krafft13}. Taking these concerns into account, magnetohydrodynamic (MHD) simulations 
cannot simulate such beam-plasma systems. Therefore, in this study we model self-consistently the large scale inhomogenieties in the background plasma, and use an emission proxy model to simulate the type \rom{3} burst. Using this approach we will demonstrate that inhomogeneities naturally give rise to the fine-structures (striations) within fundamental emission of type \rom{3} radio bursts that resemble the fine-structure detected in the latest spacecraft observation (Figure \ref{fig:type3_example}). Since it is the electrostatic non-linear interactions which ultimately give rise to the harmonic emission, this process is not modelled self-consistently here. 

Previous remote sensing studies have suggested the structuring of type \rom{3} bursts by fast mode waves~\citep{Goddard2016, Kontar2017, Kolotkov2018, Kaneda2018} which are formed by the dispersive evolution of MHD waves propagating along a non-uniform plasma~\citep{nakariakov2004fastwave}. However, studies made using high frequency and time resolution radio observations have insisted that the fine structures are formed due to coronal density fluctuations and are a multi-scale phenomena~\citep{Mugundhan2017}. The density fluctuations at large scales were reported to be responsible for the interplanetary type \rom{3} fine structures by~\cite{Jebaraj2023Type3} who also showed that the scale size of density fluctuations increase with increasing distance from the Sun. Here, the scale sizes generating the fine structures were prescribed to be 100-1000 Debye lengths, i.e., the length corresponding to the relaxation of the electron beam.

Recent studies using the FIELDS/RFS measurements have made a number of important discoveries regarding type \rom{3} bursts \citep[][]{Pulupa20, Jebaraj23b}. Of particular importance is the distinct morphology of the fundamental emission (highlighted as ``structured" in Figure~\ref{fig:type3_example}), which is strongly structured and was previously seen as an outlier among hectometer type \rom{3} bursts \citep[][]{Jebaraj2023Type3}. This finding, together with the mechanism for their generation proposed in \citep{Jebaraj2023Type3} would indicate that structured emissions are a natural consequence when density fluctuations are abundant. This leads to a qualitative assessment that density fluctuations must be present in abundance at heights corresponding to hecto-kilometer wavelengths. 

In this study, we use MHD simulations along with an emission proxy to study striations in the fundamental emission band of type \rom{3} radio bursts as a consequence of Alfv\'en wave propagation and evolution in the solar wind. Previously, \cite{Kolotkov2018} modelled such striations by fitting a modulated Newkirk density model~\citep{Newkirk1961} onto an observed type \rom{3} burst. They utilised a fitting approach to achieve agreement between the observed intensity and modelled emission intensity variations to explain the striations in type \rom{3} bursts through density modulations. In this study, we adopt a separate approach of modelling the radio burst striations through self-consistent density fluctuations that are present in the solar wind. The observed emission fluxes of such a radio burst might then similarly be a means of probing the underlying density fluctuations inside a flux tube which is the source of the observed type \rom{3} burst. To perform this study, we first initialise a solar wind with a continual injection of Alfv\'en waves which undergo PDI to generate density variations. Then, a suprathermal radiation source propagating in this quasi-steady wind is modeled to obtain the dynamic spectrum. The study finds that striations in the type \rom{3} burst are generated by the density variations in the quasi-steady wind. The manuscript describes the simulation setup in section~\ref{sec:methods}, and the modelled type \rom{3} burst in section~\ref{sec:type3-burst}.

\section{Solar wind model with PDI-driven density fluctuations}     \label{sec:methods}
The solar wind is modelled within a narrow, open magnetic flux-tube which is centered on a radial magnetic field line. The geometry of the flux-tube is parametrised by specifying the cross-sectional area ($a$) to be proportional to the flux tube expansion ($f$)~\citep{kopp1976dynamics},
\begin{equation}
 a = a_0\left(\frac{r}{r_0}\right)^2 f
\end{equation}
where $a_0$ is the cross-sectional area at the reference height $r_0$. By flux conservation the magnetic field component along the flux-tube satisfies
\begin{equation}
    B_r = B_r(r_0) \frac{a_0}{a}.
\end{equation}
The functional form providing the expansion factor $f$ is chosen as
\begin{equation}
    f = \frac{f_\mathrm{max}\exp\left(\left(r-R_1\right)/\sigma_1\right) + f_1}{\exp\left(\left(r-R_1\right)/\sigma_1\right) + 1}.
    \label{eq:f}
\end{equation}
Here $f_1$ is chosen so that $f(r_0) = 1$, and $f_\mathrm{max} = 3$, $R_1 = 1.3~R_\odot$, and $\sigma_1 = 0.5~R_\odot$ are constants with $R_\odot$ denoting the solar radius. Similar flux-tube geometries have been used in previous studies to investigate solar wind properties in open magnetic field regions~\citep{chandran2011incorporating, shoda2018self}. In particular, \cite{sishtla2022flux} discusses the suppression of the PDI for lower Alfv\'en wave frequencies by higher values of $f_\mathrm{max}$. 

\subsection{Forming the solar wind} \label{subsec:solar-wind}

The solar wind evolves in the flux-tube through the dynamical evolution of the plasma by considering a one-dimensional MHD description including the relevant physical processes of gravity and ad-hoc coronal heating,
\begin{equation}
    \frac{\partial}{\partial t}\rho + \frac{1}{a}\frac{\partial}{\partial r}\left(a\rho v_r\right) = 0
\label{eq:density}
\end{equation}
\begin{eqnarray}
    \frac{\partial}{\partial t}\left(\rho v_r\right) + \frac{1}{a}\frac{\partial}{\partial r}\left[a\left(\rho v_r^2 + p + \frac{\mathbf{B^2_\perp}}{2\mu_0}\right)\right] = \left(p + \frac{\rho \mathbf{v^2_\perp}}{2}\right)\frac{1}{a}\frac{\partial}{\partial r} a - \rho \frac{G M_\odot}{R_\odot^2}
\label{eq:MHD-vr}
\end{eqnarray}
\begin{eqnarray}
    \frac{\partial}{\partial t}\left(\rho \mathbf{v_\perp}\right) + \frac{1}{a}\frac{\partial}{\partial r}\left[a\left(\rho v_r\mathbf{v_\perp} - \frac{B_r\mathbf{B_\perp}}{\mu_0}\right)\right] = - \frac{1}{2a}\frac{\partial a}{\partial r}\left(\rho v_r\mathbf{v_\perp} - \frac{B_r\mathbf{B_\perp}}{\mu_0}\right)
\end{eqnarray}
\begin{equation}
    \frac{1}{a}\frac{\partial}{\partial r}\left(aB_r\right) = 0
\end{equation}
\begin{eqnarray}
    \frac{\partial}{\partial t} \mathbf{B_\perp} + \frac{1}{a}\frac{\partial}{\partial r}\left[a(\mathbf{B_\perp}v_r - B_r\mathbf{v_\perp}) \right] = \frac{1}{2a}\frac{\partial a}{\partial r}\left(\mathbf{B_\perp}v_r - B_r\mathbf{v_\perp}\right)
\end{eqnarray}
\begin{eqnarray}
    \frac{\partial}{\partial t} \fancyE
    + \frac{1}{a}\frac{\partial}{\partial r}
      \left[a \left(v_r \left\{ \fancyE + p + \frac{\mathbf{B^2}}{2\mu_0}- \frac{B_r^2}{\mu_0} \right\} 
                   - B_r\frac{\mathbf{B_\perp \cdot v_\perp}}{\mu_0} \right)\right]
    = -\rho gv_r + S
\label{eq:energy}
\end{eqnarray}
where,
\begin{equation}
    \fancyE = \frac{1}{2}\rho v^2 + \frac{P}{\gamma-1} + \frac{B^2}{2\mu_0}
\label{eq:def:energy}
\end{equation}
\begin{equation}
    S = S_0 \mathrm{exp}\left(-\frac{r}{L}\right).
\label{eq:def:exp-heating}
\end{equation}
The quantities $\rho$, $\mathbf{v}$, $\mathbf{B}$, $\fancyE$, and $p$ correspond to the mass density, plasma bulk velocity, magnetic field, total energy density, and thermal pressure, respectively. The directions along and transverse to the flux tube are denoted by $r$ and $\perp$, respectively. 
We use an ideal-gas equation of state and set the adiabatic index to be $\gamma = 5/3$. To obtain a steady-state solar wind that approximates a Parker-like outflow, we incorporate an additional energy source term in Equation~\ref{eq:def:exp-heating}~\citep{pomoell2015modelling, mikic2018predicting} with $S_0 = 0.5\times 10^{-6}$~Wm$^{-3}$ and $L = 0.4~R_\odot$~m. 

The MHD equations~\ref{eq:density}-\ref{eq:energy} are evolved in time using the strong stability preserving (SSP) Runge-Kutta method to advance the semi discretised equations in time. The simulation domain is discretized using $3000$ cells spaced logarithmically from the solar corona (lower boundary). The method employs the Harten–Lax–van Leer (HLL) approximate Riemann solver supplied by piece-wise, linear slope limited interface states. The numerical method also employs the constrained transport technique to ensure the magnetic field solution is divergence free~\citep{Kissmann2012}. These methods have been used applied in previous studies of the solar corona~\citep{pomoell2012influence, Sishtla23}. The simulation domain extends radially from $r = 1.03~R_\odot$ to $r = 30~R_\odot$, with the $\theta,~\phi$ directions being transverse to the background magnetic field. At the low-coronal inner boundary, 
the mass density $\rho$, temperature $T$, and radial magnetic field $B_r$ are fixed to the values $\rho_0 = 8.5\times 10^{-13}~\mathrm{kg}~\mathrm{m}^{-3}, ~T_0 = 8\times 10^5~\mathrm{K}, ~B_0 = 2.5 \, \mathrm{G}$.

\subsection{Development of PDI in the solar wind}\label{subsec:PDI}
After a quasi-steady solar wind configuration is obtained by integrating the MHD equations in time, we continually introduce monochromatic Alfv\'enic perturbations at the lower boundary ($r = 1.03~R_\odot$) by perturbing the transverse components of the magnetic field. These Alfv\'enic fluctuations are described using Els\"asser variables~\citep{elsasser1950hydromagnetic}, and denote correlations between fluctuating velocity and magnetic field as $\mathbf{z_{\theta, \phi}}^\pm = \mathbf{v_{\theta, \phi}}^\pm \pm {\mathbf{B_{\theta, \phi}}}/{\sqrt{\mu_0\rho}}$. In particular, at the inner radial boundary we impose
\begin{equation}
	\mathbf{z}^- = \mathrm{Z_0}\sqrt{2}\sin(2\pi f_0 t) \mathbf{\hat{\theta}} + \mathrm{Z_0}\sqrt{2}\cos(2\pi f_0 t) \mathbf{\hat{\phi}}
\end{equation}
where $\mathrm{Z_0} = 32$ km s$^{-1}$ and $f_0 = 1~$mHz. This corresponds to injecting an anti-sunward (outward) circularly polarised Alfv\'en wave into the magnetic flux-tube. It is known that the MHD equations permit the existence of finite-amplitude Alfv\'en waves with circular polarisation which do not generate variations in the total magnetic field and thermal pressure as they propagate~\citep{barnes1974large, hollweg1974transverse}. 
In low $\beta$ plasma 
the circularly polarised wave can couple with small random fluctuations in the magnetic field, such as when the background field has gradients, to trigger the parametric decay instability~\citep{goldstein1978instability}. This instability occurs in our simulation as the solar wind obtained through Section~\ref{subsec:solar-wind} contains $\beta < 0.45$ in the simulation domain prior to the injection of the Alfv\'en wave. The specific choice of $\mathrm{Z_0}$ and $f_0$ ensures that the injected Alfv\'en wave does not significantly drive the solar wind as the injected energy is much smaller than that required to sustain the wind in an open field region~\citep{Withbroe1977}, while still ensuring a sufficient growth rate for the PDI. The PDI results into decay of the anti-sunward propagating Alfv\'en wave decays into a reflected sunward propagating Alfv\'en wave, and anti-sunward propagating MHD slow wave. Therefore, the PDI satisfies the conditions $\mathbf{k}_1 = \mathbf{k}_2 + \mathbf{k}_3$ and $\omega_1 = \omega_2 + \omega_3$ where the subscripts 1-3 refer to the anti-sunward and sunward Alfv\'en waves, and the slow wave, respectively~\citep{chandran2018parametric}. 
The development of the PDI instability depends on the frequency of the injected Alfv\'en wave~\citep{shoda2018frequency}. In addition, the expansion factor of the flux-tube suppresses the development of PDI for lower frequency Alfv\'en waves~\citep{sishtla2022flux}. In this study we have chosen appropriate values of $f_\mathrm{max}$, and $f_0$ which excite the PDI instability, with \cite{sishtla2022flux} providing a complete description of the Alfv\'en wave dynamics inside a flux-tube as prescribed in this simulation.

\begin{figure*}[ht]
    \centering
    \includegraphics[width=1.0\linewidth]{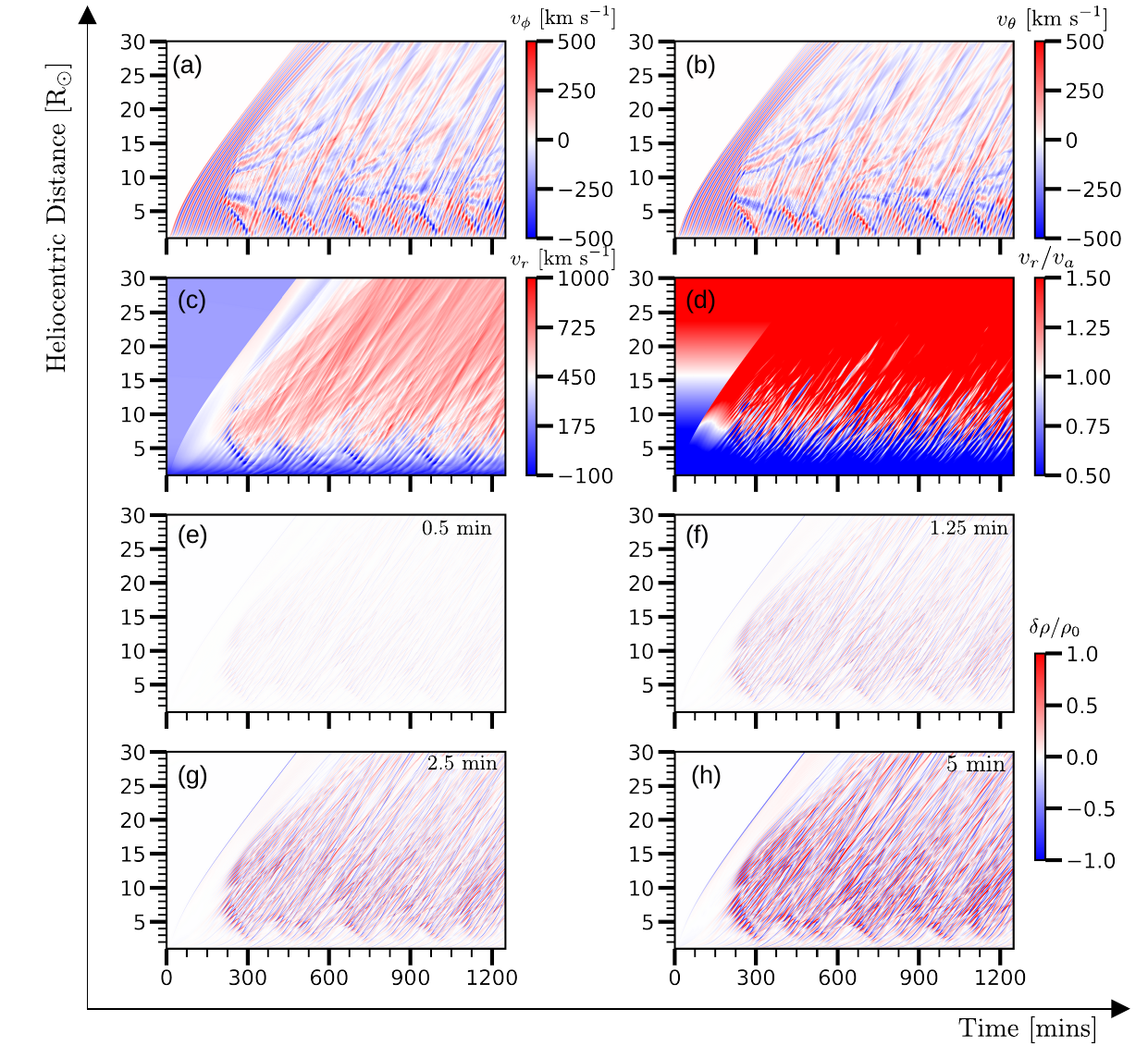}
    \caption{Dynamical evolution of the solar wind plasma along the flux-tube. In every panel, each column represents the quantity (indicated with the colorbar) at the given time along the flux tube. The panels show the parameters (a) $v_\phi$, (b) $v_\theta$, (c) $v_r$, and (d) $v_r/v_a$ for the case where a $1~$mHz Alfv\'en wave is continually injected starting at $t = 0$. In addition, the density fluctuations defined as $\delta\rho/\rho_0$ where $\rho_0$ is the time averaged density are presented for different averaging intervals of (e) $0.5~$mins, (f) $1.25~$mins, (g) $2.5~$mins, and (h) $5~$mins.}
    \label{fig:solar_wind}
\end{figure*}
In Figure~\ref{fig:solar_wind}(a)-(d) the dynamical evolution of the solar wind is presented after the start of the injection of the Alfv\'en wave. Panels (a) and (b) show the transverse ${v_\phi}$ and ${v_\theta}$ velocity components that are perturbed due to the introduction of the circularly polarised wave. The start of the wave injection is at $t=0$~mins ($x$-axis) and the waves are seen to propagate through the corona towards the outer boundary. From $t \approx 225~$ mins onwards, a significant change in the dynamics originating at approximately between $r \sim 6 R_\odot$ and $r \sim 8 R_\odot$ occurs. At this time, the creation of two different space-time paths in $v_{\phi,\theta}$ is visible, with one path propagating sunward and the other anti-sunward. At this same instance, the development of fluctuations in $v_r$ (panel (c)), along with variations in the Alfv\'en critical point (panel (d)) is evident. 

In panels (e)-(h) the evolution of the level of density fluctuations is shown, defined as $\delta\rho/\rho_0$ where $\rho_0$ is the time-averaged density, and $\delta\rho = \rho - \rho_0$ is the corresponding density fluctuation around $\rho_0$. Panels (a)-(d) present the density fluctuations computed using different averaging windows, which allows us to study the fluctuations at different frequencies. This figure shows that the density fluctuations start to develop at various scales at $t\approx 225~$ min, with higher absolute values of $\delta\rho/\rho_0$ in panel (d) where the averaging interval captures frequencies closer to the pump wave. It can then be seen through the emerging density fluctuations in panels (e)-(h) along with the decay of the injected wave at the same time in panels (a)-(d) that it is the parametric decay instability that is excited at $t\approx 225~$ min. 

The different tracks in panels (a), (b) correspond to the waves propagating with phase speeds $v_r\pm v_a$ (where $v_a$ is the Alfv\'en speed) in the anti-sunward and sunward directions~\citep{Verdini2009}. This modified $\mathbf{v_\perp}$ then drive variations in $v_r$ through Equation~\ref{eq:MHD-vr} which we observe in panel (c). Moreover, the $v_{\phi,\theta}$ tracks indicate a quasi-periodic structure of the solar wind as the Alfv\'en waves are continually injected. This structure indicates that the density fluctuations would be similarly quasi-period which is seen in panels (e)-(h).

\section{Modelling the effect of PDI on the morphology of type \rom{3} radio bursts} \label{sec:type3-burst}
Figure \ref{fig:type3_example} shows two example intervals during PSP's close encounter 9. The most striking feature in the figure is the strongly structured morphology of the type \rom{3} bursts that are uniquely observed within the fundamental emission. Assuming that the fundamental emission is generated via linear mechanisms \citep[][]{Voshchep15a, Voshchep15b, Krasnoselskikh19, Jebaraj2023Type3}, the generation of morphological features within a type \rom{3} should be facilitated by density inhomogeneities at different spatial scales present in the medium. To this degree, we model the generation of the fundamental emission as a near-relativistic propagating source that interacts with the ambient plasma to generate electromagnetic (EM) waves based on the local electron plasma frequency $f_\mathrm{pe}$, 
\begin{equation}
    f_\mathrm{pe} \, \mathrm{[MHz]} \approx 9\times 10^{-3}\sqrt{n\, \mathrm{[cm^{-3}]}}
\label{eq:fpe}
\end{equation}
where $n = \rho/m_p$ is the number density and $m_p$ is the proton mass. Since it is not possible to simulate beam-plasma interactions within a MHD plasma description, we construct an emission proxy approximating this process for a given speed $v_e$ and time of injection $t_i$ of the radiation source. The radiation source would represent a sampling of the suprathermal electron beam propagating along the open magnetic field lines. If we assume a constant velocity for the radiation source, it arrives at the outer boundary at time $t = d/v_e$, with $d$ representing the length of the flux tube in the MHD simulation. Finally, the location of the radiation source $r_e$ at any intermediate time when it is in the simulation domain can be expressed as $r_e(t) = r_0 + v_e t$, where $r_0$ is the location where the radiation source originates. Thus, we can sample the location of the radiation source as it propagates and use the background density from the MHD simulation to calculate the EM emission at the location $r_e$ using Equation~\ref{eq:fpe}. In other words, we make a snapshot of the background plasma which represents the variations in the background electron plasma frequency in the modelled flux tube.
\begin{figure*}
    \centering
    \includegraphics[width=\linewidth]{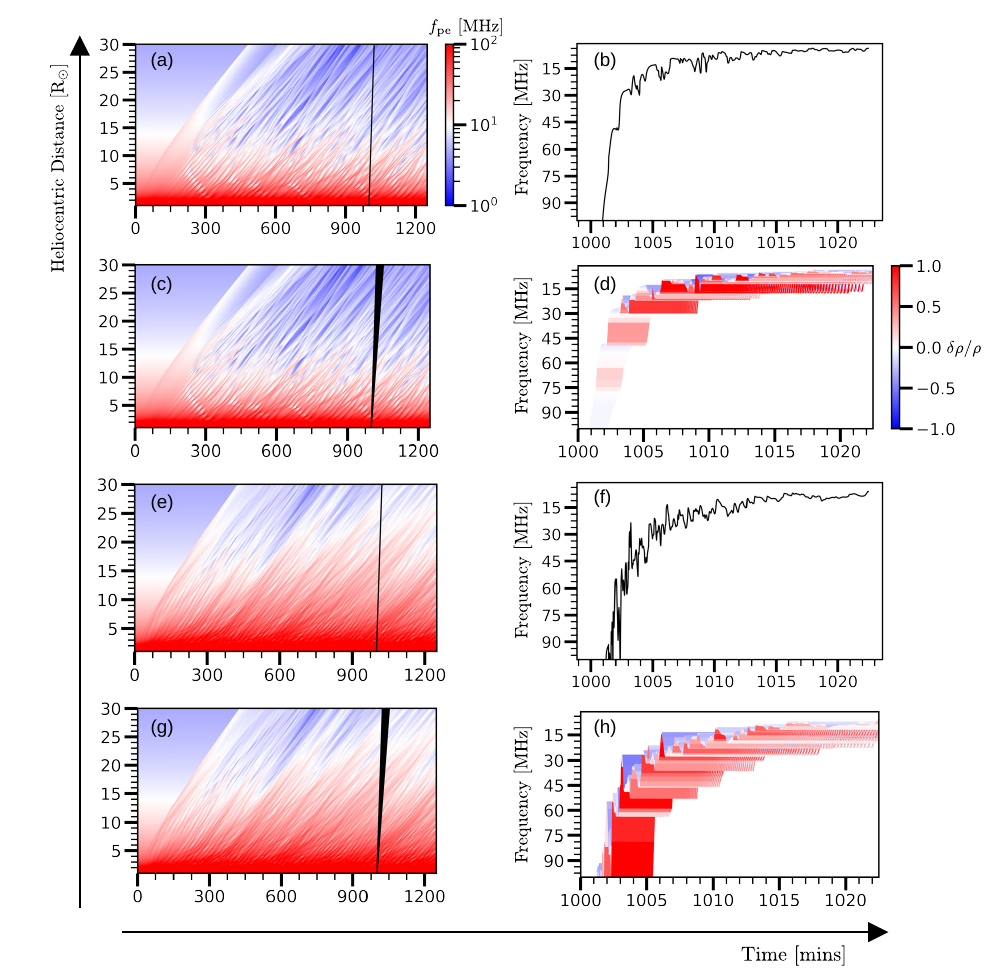}
    \caption{The spatio-temporal dependence of the (a) local plasma frequency in the solar wind, and (b) the associated simulated dynamic spectrum for a radiation source injected at $t=1000~$mins is shown. The path of the radiation source is annotated in (a). Subsequently, panels (c) and (d) present the local plasma frequency annotated with the path of a beamlet 
    and the associated dynamic spectrum. In panel (d) the dynamic spectrum is coloured based on the $5~$mins averaged density fluctuations encountered by the beamlet. Similar to panels (a) and (b), the panels (e) and (f) present the modelled emission for a radiation source but in a solar wind injected with Alfv\'en waves with frequencies $1,~2,~$and $3~$mHz. The panels (g) and (h) present the modelled emissions for a similar beamlet as in panels (c) and (d) but in a solar wind injected with the Alfv\'en waves of three different frequencies.}
    \label{fig:type3_burst}
\end{figure*}

This process of simulating the Type \rom{3} radio burst is illustrated in Figure~\ref{fig:type3_burst}. In panel (a), we show the solar wind's local electron plasma frequency, which contains quasi-periodic variations as the PDI generates density fluctuations. We then inject a fast moving radiation source at $t = 1000~$min with a velocity of $0.05c$, which propagates through the plasma along the trajectory annotated in panel (a). We sample the radiation source's location as it propagates and plot the associated emission frequency in panel (b), representing the snapshot of the plasma along the radiation source propagation path. This simulated dynamic spectrum in panel (b) is plotted for emission frequencies between $1$ MHz and $100$ MHz, after the radiation source propagates away from the low solar corona which is associated with significantly higher frequencies due to the increased density (Equation~\ref{eq:fpe}). The striations of various scales and amplitudes caused by density fluctuations in panel (b) is characteristic of a Type \rom{3} radio burst. As this methodology does not involve simulating beam-plasma interactions, the features of the simulated dynamic spectrum are not affected by the exact value of the near-relativistic velocity of the radiation source.

We can now extend this methodology to model a beamlet, a collection of radiation sources with an associated beam angle, as illustrated in panels (c) and (d). The modelled beamlet consists of multiple radiation sources injected at the same time but with slightly varying constant speeds along the radial magnetic field. We define a beam angle in phase-space, thus denoting the angle between the fastest and slowest radiation source. In panels (c) and (d) we launch $100$ radiation sources at $t = 1000~$mins with the fastest speed being $0.05c$ with a beam angle of $5\times 10^{-6}$ degrees. The local electron plasma frequency as calculated from Equation~\ref{eq:fpe} along with the path of the beamlet is presented in panel (c). The simulated dynamic spectrum due to this beamlet is shown in panel (d) 
color-coded by the $5~$min averaged density fluctuations $\delta\rho/\rho_0$ at the location of emission.
These $5~$min averaged fluctuations are the larger scale density perturbations present in the simulated solar wind. The density fluctuations at such scales have greater $\left|\delta\rho/\rho_0 \right|$ as they are closer to the injected Alfv\'en wave frequency due to the decay process involving the pump wave parametrically generating lower frequency wave distributions~\citep[see e.g.][]{chandran2018parametric}. The dynamic spectrum in panel (d) is seen to rise with an accumulations of plasma ($\delta\rho > 0$) and falls due to a deficit ($\delta\rho < 0$) as the emission frequency of the electron beamlet is higher for an increased density (Equation~\ref{eq:fpe}). Therefore, the striations in the dynamic spectrum are a consequence of a time of flight effect of the electrons propagating in the wind.

\subsection{Spectral striations are the natural response to beam propagation in inhomogeneous plasma}
Consider a radiation source propagating through the coronal plasma, leading to emissions at the plasma frequency,
\begin{equation}
    2\pi f_\mathrm{pe} = \sqrt{\frac{n e^2}{\epsilon_0 m_e}}
\end{equation}
where $n$ is the electron number density, $e$ is the elementary charge, $m_e$ is the mass of the electron, and $\epsilon_0$ is the permittivity of vacuum. Then, for any perturbations in $n$, the corresponding perturbation in emission frequency is
\begin{equation}
    2\pi \delta f_\mathrm{pe} = \frac{1}{2}\sqrt{\frac{e^2}{n\epsilon_0 m_e}}\delta n = \frac{\delta n}{2n}\sqrt{\frac{ne^2}{\epsilon_0 m_e}} = \frac{\delta n}{2n} 2\pi f_\mathrm{pe}
\end{equation}
Therefore,
\begin{equation}
    \frac{\delta f_\mathrm{pe}}{f_\mathrm{pe}} = \frac{1}{2}\frac{\delta n}{n} = \frac{1}{2}\frac{\delta \rho}{\rho}
\label{eq:freq-perturb}
\end{equation}
For cases when $-1 \leq \delta\rho/\rho \leq 1$, then $-f_\mathrm{pe}/2 \leq \delta f_\mathrm{pe} \leq f_\mathrm{pe}/2$. Thus, increased $\left|\delta\rho/\rho \right|$ will directly correlate with greater striations in the dynamic spectrum. To illustrate this further we present Figures~\ref{fig:type3_burst}(e),~\ref{fig:type3_burst}(f) and Figures~\ref{fig:type3_burst}(g),~\ref{fig:type3_burst}(h), which show the simulated dynamic spectrum for both a radiation source and a beamlet propagating in a solar wind which is subject to the simultaneous injection of Alfv\'en waves at the three frequencies $1$ mHz, $2$ mHz, and $3$ mHz. This is in contrast to the preceding sections where $\delta\rho/\rho_0$ variations were the consequence of a single frequency wave injection. By introducing additional wave modes we expect more wave-wave interactions leading to greater number of striations. 

Panel (e) presents the dynamic spectrum for a radiation source injected at $t~=~1000~$mins. As compared to the case of injecting a single wave mode in panel (a), panel (e) shows finer fluctuations in the local plasma frequency, and this is captured along the electron path in panel (f). If we further launch $100$ radiation sources at $t~=~1000~$mins with a beam angle of $5\times 10^{-6}$ degrees, the simulated dynamic spectrum in panels (g), (h) similarly shows greater striations than panels (c), (d). Our model therefore suggests that the striations in the dynamic spectrum are a consequence of a time of flight effect of the electrons propagating in the wind that embeds density perturbations. Thus, in a more general solar wind plasma containing a well-developed spectrum of density fluctuations, the time of flight effect of the propagating electrons would lead to structured striations in the dynamic spectra of type \rom{3} bursts such as those shown in Figure~\ref{fig:type3_example}.

\section{Conclusion}     \label{sec:conclusion}
In this study we simulated the effect of coronal density fluctuations on the fundamental band of a type \rom{3} radio burst. The fluctuations were self-consistently generated by injecting an Alfv\'en wave into the low corona, which subsequently propagates into the higher corona and triggers the PDI instability via which density fluctuations are generated in the solar wind. Such instabilities and the associated density fluctuations are expected to be present in the solar corona where the low plasma beta conditions result in a sufficient growth rate for PDI. The synthetic type \rom{3} burst produced in this study was affected by the multi-scale density fluctuations in the solar wind. These scales correspond to hundreds to thousands of Debye length with a maximum of the relaxation length of the beam. In this regard, the synthetic type \rom{3} represent a snapshot of the fluctuations of the same level in the background density along a flux tube. This is relevant considering recent PSP observations indicating that most type \rom{3} bursts recorded during close encounters are structured at various spatio-temporal scales \citep[][]{Pulupa20, Jebaraj23b}. Furthermore, the presence of dynamic wave activity in the inner heliosphere has also been widely recognized~\citep{Malaspina_2020, Mozer2023}. It is possible that the consistent appearance of structured type \rom{3} radio bursts observed when the probe was close to the Sun indicates the presence of a fundamental process which mediates this. The results presented in this study indicate the possibility of radio bursts affected by Alfv\'enic turbulence in the solar corona which might lead them to exhibit variations in the emission frequency based on the spectral properties of turbulent density fluctuations. Conversely, examining the emissions in the radio burst might provide a means of probing the density fluctuations inside the flux-tube where the type \rom{3} burst originated. The exact spatial scales of the fluctuations which affect the beam-plasma system are highly dependent upon the characteristics of the beam such as its density. As previously mentioned, this relaxation length of the electron beam will then form the upper limit for the observed striations \citep[][]{Jebaraj2023Type3}.

While our simulations self-consistently generate density fluctuations via the PDI, there are also other processes close to the Sun that may generate density fluctuations. Inhomogenous plasma conditions in the direction transverse to the propagating wave vector lead to phase mixing which is an alternate source of generating fast waves, and by extension compressive fluctuations~\citep{nakariakov2000nonlinear}. This phase mixing can independently generate turbulent structures in the solar wind~\citep{magyar2017generalized}. More recently, it has been seen that small-scale reconnection on the solar surface can drive jetlets and switchbacks which contribute to the compressive nature of the wind~\citep{Raouafi2014, Raouafi2023}. This is well complemented by \cite{Bale19, Bale23} who have found a strong correspondence between the solar wind arising from coronal hole boundaries and reconnection driven processes. Therefore, compressive fluctuations driven by recconection may also contribute to the turbulent environment which can affect the morphology of the type \rom{3} bursts.
Thus, density fluctuations that are responsible for causing variations in the emission frequencies of radio sources may arise from a multitude of non-linear processes in the solar corona and further out in interplanetary space. Nevertheless, our results show that even a simple scenario where PDI is responsible for producing density variations in the solar wind is able to produce synthetic radio spectra consistent with the observed emission. Subsequently, analysing type \rom{3} bursts in a high temporal resolution, such as through PSP, can provide indirect information on the relative density fluctuations~\citep{Krupar2020} present in the wind. The scenario presented in this study proves as a first step in utilizing remote sensing observations such as the ones from radio wavelengths, e.g. using a forward modelling approach, to constrain the evolution of plasma in regions not generally amenable to direct measurements. 


\section*{Acknowledgements}
Parker Solar Probe was designed, built, and is now operated by the Johns Hopkins Applied Physics Laboratory as part of NASA’s Living with a Star (LWS) program (contract NNN06AA01C). Support from the LWS management and technical team has played a critical role in the success of the Parker Solar Probe mission.
The work has been supported by the Finnish Centre of Excellence in Research on Sustainable Space (FORESAIL; grant no. 336807). This is a project under the Academy of Finland, and this research has been supported by the European Research Council (SolMAG; grant no. 724391) as well as Academy of Finland project SWATCH (343581). Open access is funded by Helsinki University Library. I.C.J, M.P, and S.D.B acknowledge support from the International Space Science Institute (ISSI) in Bern, through ISSI International Team project No.557, “Beam-Plasma Interaction in the Solar Wind and the Generation of Type \rom{3} Radio Bursts”. I.C.J also acknowledges support from the ISSI visiting scientist program. The authors would also like to thank Dr. V. V. Krasnoselskikh for the active discussions on the topic of parametric decay instability and radio generation mechanisms. 

\bibliographystyle{aasjournal} 
\bibliography{example}      

\end{document}